\documentclass[twocolumn,showpacs,preprintnumbers,amsmath,amssymb,a4paper,superscriptaddress]{revtex4}

\usepackage{amsmath}
\usepackage{graphicx}
\usepackage{dcolumn}
\usepackage{bm}

\newcommand {\nc} {\newcommand}
\nc {\ve} [1] {\mbox{\boldmath $#1$}} \nc {\la} {\mbox{$\langle$}}
\nc {\ra} {\mbox{$\rangle$}} \nc {\beq} {\begin{eqnarray}}
\nc
{\eeqn} [1] {\label{#1} \end{eqnarray}}

\nc {\eol} {\nonumber \\} \nc {\half} {\mbox{$\frac{1}{2}$}}

\begin{document}

\title{Analytical treatment of bosonic d-wave scattering in isotropic harmonic waveguides}
\author{P. Giannakeas}
\email{pgiannak@physnet.uni-hamburg.de}
\affiliation{Zentrum f\"{u}r Optische Quantentechnologien, Universit\"{a}t Hamburg, Luruper Chaussee 149, 22761 Hamburg,
Germany,}

\author{V.S. Melezhik}
\email{melezhik@theor.jinr.ru}
\affiliation{Bogoliubov Laboratory of Theoretical Physics, Joint Institute for Nuclear Research,
Dubna, Moscow Region 141980, Russian Federation,}

\author{P. Schmelcher}
\email{pschmelc@physnet.uni-hamburg.de}
\affiliation{Zentrum f\"{u}r Optische Quantentechnologien, Universit\"{a}t Hamburg, Luruper Chaussee 149, 22761 Hamburg,
Germany,}

\date{\today}

\begin{abstract}
We analyze d-wave resonances in atom-atom scattering in the presence of harmonic confinement by employing a higher partial wave pseudopotential. 
Analytical results for the scattering amplitude and transmission are obtained and compared to corresponding numerical ones,
which employ the Lennard-Jones potential.
Qualitative agreement is observed for weak confinement.
For strong confinement the pseudopotential does not capture the s- and d-wave interference phenomena yielding an asymmetric Fano profile for the transmission resonance.

\end{abstract}


\maketitle

\section{Introduction}

The characteristics of ultracold bosonic two-body scattering processes provide a key ingredient for modeling and understanding of the properties of degenerate atomic gases.
The dominant s-wave interactions of the two-body collisions can be conveniently modeled by Huang-Fermi's pseudopotential \cite{fermi34, huang57, huang63}.
Over the last decade, this pseudopotential has become an
essential theoretical tool in particular for analytical studies of ultracold
systems, e.g., see Busch et. al. \cite{busch98} for two atoms in harmonic traps, Olshanii for
atomic collisions in cylindrical waveguides \cite{olshanii98},
where the so-called confinement-induced
resonances (CIRs) was predicted and experimentally observed in \cite{kinoshita04,paredes04,haller09}, or in refs. \cite{tiesinga00, blume02, bolda02} which employed approaches based on
the energy dependent scattering lengths, $a_s(E)$.

The theory of pseudopotentials has been developed further in order to describe scattering processes involving higher-partial waves, i.e., with relative angular momentum $\ell > 0$ \cite{huang57, huang63, derevianko05, stock05, idziaszek06, idziaszek09, stampfer08, pricoupenko06,huang89, demkov81, frolov03}.
Higher-partial wave interactions represent an intriguing perspective in research on ultracold gases and are expected to provide novel many-body phenomena, such as unconventional superfluidity and superconductivity \cite{rey09, deb09, hofstetter02}. 
So far, theoretical studies with pseudopotentials for $\ell>0$ have mainly been used to describe fermionic systems, e.g., in quasi-one-dimensional \cite{stock05, idziaszek09, kanjilal04} or quasi-two-dimensional \cite{idziaszek06,pricoupenko06} geometries, which have been studied experimentally in \cite{guenter05}.
For the case of bosons an analytical description of collisions in a planar waveguide has been derived, where the cross-section possesses s- and d-wave resonances \cite{pricoupenko06}.
In harmonic waveguides higher-partial wave interactions were shown to have important impact on bosonic scattering \cite{gian11} through the interference of d-wave resonances (DWRs) with the underlying continuum.
A theoretical treatment of d-wave scattering of bosons in the case of quasi-1D geometries is, however, still needed.

In this work we derive an approximative analytical solution for bosonic collisions in a harmonic waveguide, which goes beyond mere s-wave scattering.
Our starting-point is to assume that the transversal confinement frequency is weak enough in order to avoid a strong interference between the free space d-wave shape resonance and the s-wave scattering. 
The latter permits to model the interatomic potential through a zero-range regularized d-wave pseudopotential suggested in \cite{idziaszek06, idziaszek09}, 
which is parametrized by a single parameter, the energy dependent d-wave scattering length ($a_d(k)$). The associated regularization operator contains exclusively derivatives. 
Taking into account the virtual excitations occurring during the scattering process we analytically describe the appearance of the DWR.
An important aspect of our approximative formula for the d-wave scattering amplitude is that one can map the total Hamiltonian onto an effective one-dimensional Hamiltonian, 
where the interaction is modeled by a one-dimensional contact potential. 
The strength of the contact potential depends on the above-mentioned d-wave scattering amplitude, similar to the case of the s-wave CIR \cite{olshanii98}.

In order to verify the validity of the approximative analytical results we compare them with corresponding numerical calculations, where the interatomic interaction is modeled by a Lennard-Jones potential.  
A good agreement is obtained only for a sufficiently weak confinement. 
For strong confinement the pseudopotential does not capture the s- and d-wave interference and 
therefore lacks the description of the numerically obtained Fano asymmetric lineshape of the transmission coefficient \cite{fano61}.

The sectioning of this paper is as follows. 
In Sec. II we review the recent developments of pseudopotential theory for higher partial wave interactions. 
In Sec. III we describe and discuss the d-wave scattering process. 
Sec. IV contains a comparison of the analytical results and the numerical calculations and Sec. V provides a brief summary and conclusions.

\section{Pseudopotentials for higher partial wave interactions}

The pseudopotential approximation to the interatomic interaction is an ubiquitous tool for the theoretical description of ultracold gases: the atomic interactions are modeled in terms of a contact potential.
Huang and Yang \cite{huang57,huang63} derived a generalized pseudopotential for all partial waves, which represents a multipolar expansion over
delta function contributions. The Huang-Yang approach (HY) was revised by Derevianko \cite{derevianko05}, who corrected an algebraic error for $\ell>0$.
However, other derivations have been developed for higher partial wave pseudopotentials, which differ from the HY-approach, by using a more rigorous treatment of the derivatives of the delta-functions
in the framework of distribution theory \cite{idziaszek06,idziaszek09} yielding the following representation:

\begin{align}\nonumber
V_{ps}(\mathbf r) \Psi(\mathbf r)&=\sum_{\ell=0}^\infty\dfrac{(-1)^{\ell} \left[(2\ell+1)!!\right]^2}{4\pi(2\ell)!\ell!}g_{\ell}\frac{\delta^{(\ell)}(r)}{r^2}\\
&\times\left[\dfrac{\partial^{2\ell+1}}{\partial {r'}^{2\ell+1}} {r'}^{\ell+1} \int d \mathbf{\Omega'} P_{\ell} (\hat{\mathbf{r}} \cdotp \hat{\mathbf{r}}') \Psi(\mathbf r')\right]_{\mathbf{r'}=0},
\label{1}
\end{align}

\noindent where $g_{\ell}=\hbar^{2} a_{\ell}^{2\ell+1}(k)/(2\mu)$, with $a_{\ell}^{2\ell+1}(k)=-\text{tan}\delta_{\ell}(k)/k^{2\ell+1}$ defined as the $\ell$-th energy-dependent scattering length \cite{blume02,bolda02}.
Here, $\delta^{(\ell)}(r)$ is the $\ell$-th derivative of the delta function and
the integral over the solid angle $\mathbf{\Omega'}$ acts as a projection operator of the total wavefunction on each $\ell-$state. 
For $\ell=0$ one obtains from Eq.(\ref{1}) the Huang-Fermi pseudopotential.
However, the Eq.(\ref{1}) for $\ell>0$ differs by a factor $(2\ell+1)/(\ell+1)$ from Huang and Yang's result \cite{huang63}, whereas if we substitute in Eq.(\ref{1}) the relations
$\delta(\mathbf{r})=\delta(r)/(4\pi r^2)$ and $\delta^{(\ell)} (r)=(-1)^{\ell}\ell!\delta(r)/r^\ell$ we obtain the pseudopotential of Derevianko \cite{derevianko05}.

Additionally, Stampfer and Wagner \cite{stampfer08} introduce the following pseudopotential for higher partial waves:

\begin{align}\nonumber
\widetilde{V}_{ps}(\mathbf r)\Psi(\mathbf r)&= \sum_{\ell=0}^\infty \sum_{m=-\ell}^\ell \dfrac{ 4\pi (-1)^\ell}{(2\ell)!!}g_{\ell} Y_{\ell m}(\partial) \delta(\mathbf{r})\\
&\times\left[\dfrac{\partial^{2\ell+1}}{\partial {r'}^{2\ell+1}} {r'}^{\ell+1} \int d \mathbf{\Omega'} Y^{\ast}_{\ell m} (\hat{\mathbf{r}} ') \Psi(\mathbf r')\right]_{\mathbf{r'}=0},
\label{2}
\end{align}
where the operator $Y_{\ell m}(\partial)$, introduced by Maxwell \cite{macrobert48}, is obtained from the spherical harmonic polynomial $r^{\ell}Y_{\ell m}(\hat{\mathbf{r}})$, by replacing the Cartesian coordinates $x_{k}$
with the partial derivatives $\partial_{x_k}$. Note that Eq.(\ref{2}) is equal to Eq.(\ref{1}) as it was proven in \cite{idziaszek09} by using the relation

\begin{equation}
\frac{\delta^{(\ell)}(r)}{r^2}Y_{\ell m}(\hat{\mathbf{r}})=\frac{4\pi \ell!}{(2\ell+1)!!}Y_{\ell m}(\partial)\delta(\mathbf{r}).
\label{3}
\end{equation}

The above-mentioned pseudopotentials can be simplified \cite{idziaszek09} in order to describe interactions of a single partial wave character.
This simplification is based on the fact that in general the atomic interaction is dominated by the $\ell$-partial wave scattering in the vicinity of the corresponding $\ell$-partial wave resonance.
Thus, one can focus on a single partial wave, which yields the following pseudopotential:

\begin{eqnarray}\nonumber
V_{ps, \ell}(\mathbf r)\Psi(\mathbf r)&=& \sum_{m=-\ell}^\ell \dfrac{16\pi^{2} (-1)^\ell}{(2\ell+1)!}g_{\ell} Y_{\ell m}(\partial) \delta(\mathbf{r})\\
&&\times\left[\dfrac{\partial^{2\ell+1}}{\partial {r'}^{2\ell+1}} {r'}^{2\ell+1}  Y^{\ast}_{\ell m} (\partial ') \Psi(\mathbf r')\right]_{\mathbf{r'}=0}.
\label{4}
\end{eqnarray}

Furthermore, the sum w.r.t. $m$ can be simplified with the help of the following summation formula:

\begin{equation}
\frac{4\pi}{2\ell+1}\sum_{m=-\ell}^{m=\ell}Y_{\ell m}(\partial)Y^{\ast}_{\ell m}(\partial ')=\sum_{k=0}^{[\ell /2]} c_{k}\left(\bm{\nabla} \cdot \bm{\nabla} '\right)^{\ell-2k} \nabla^{2k} (\nabla ')^{2k},
\label{5}
\end{equation}

\noindent where $[\ell/2]$ is the largest, smaller or equal integer of $\ell/2$ and $c_{k}=\left[(-1)^{\ell}(2\ell+1)!\right]/\left[2^{\ell}k!(\ell-k)!(\ell-2k)!\right]$ are the coefficients
of the Legendre polynomials $P_{\ell}(x)=\sum_{k=0}^{[\ell/2]}c_{k}x^{\ell-2k}$.

Eqs.(\ref{4}) and (\ref{5}) constitute pseudopotentials for any $\ell$-partial wave. These pseudopotentials are expressed in terms of differential operators
which are more convenient in analytical derivations than the pseudopotential with the projection operator (see Eqs.(\ref{1}) and (\ref{2})).

\section{Quasi-1D scattering with s- and d-wave interactions}
\subsection{s- and d-wave pseudopotentials}
In the following we consider ultracold collisions of identical bosons in the presence of an external harmonic confinement.
The scattering process is taking place in the low energy regime, $ka_{\perp}\ll1$, where $k$ is the relative momentum in the unconfined degree of freedom and
$a_{\perp}=\sqrt{\frac{\mu \omega_{\perp}}{\hbar}}$ is the oscillator length defined via the trap frequency $\omega_\perp$ and the reduced mass $\mu$ of the colliding bosonic pair.
Furthermore, in order to go beyond s-wave physics we assume that the momentum $k$ is very small but not equal to zero.
Consequently, as it was shown in \cite{gian11} except for the well known s-wave CIR \cite{olshanii98}, an additional resonance emerges in this system, the so-called DWR,
which is based on the interference of a free space d-wave shape resonance with s-wave scattering.
The coexistence of s- and d-wave resonances implies interactions between the bosons of s- and d-wave symmetries (see Fig.1(c) in Ref.\cite{gian11}).

The main goal of this paper is to present an analytical model for the DWR effect in a harmonic waveguide.
However, the modeling of s- and d-wave interactions with pseudopotentials in systems of non-spherical symmetry, as the above-mentioned one, is not a trivial task.
As the symmetry of the system is cylindrical the angular momentum is not conserved and consequently, the application of pseudopotentials, Eqs. (\ref{1}) and (\ref{2}), leads to an explicit integration over all
$\ell$-partial waves.

Though, as it was proposed in \cite{idziaszek09} one can avoid such difficulties by considering a single $\ell$-partial wave in the vicinity of the $\ell$-wave resonance, where contributions from other partial waves are
sufficiently suppressed.
The latter means that in the vicinity of the s-wave resonance, we assume that the interaction can be modeled by the following pseudopotential with the help of Eqs. (\ref{4}) and (\ref{5}):

\begin{equation}
 V_{ps,s}(\mathbf{r})=\frac{2\pi \hbar^2 a_{s}(k)}{\mu} \delta(\mathbf{r}) \frac{\partial}{\partial r}[r \cdot~],
\label{6}
\end{equation}

\noindent which is the Huang-Fermi pseudopotential, with the  s-wave scattering length $a_{s}(k)$ being energy dependent. 
Note that Eq.(\ref{6}) coincides with the pseudopotential proposed in \cite{bolda02}.

Equivalently, in the vicinity of the d-wave resonance, we assume that the interaction can be modeled by the following pseudopotential \cite{idziaszek06,idziaszek09}:

\begin{equation}
 V_{ps,d}(\mathbf{r})=\frac{\pi \hbar^2 a_{d}^{5}(k)}{8\mu}\sum_{i,j,k,l} D_{ijkl}~^\leftarrow \!~(\partial_{x_{i}x_{j}}^{2}) \delta (\mathbf{r}) \frac{\partial ^5}{\partial r^5}r^{5}(\partial_{x_{k}x_{l}}^{2})  ^{\rightarrow},
\label{7}
\end{equation}

\noindent where $D_{ijkl}=\delta_{ik}\delta_{jl}-\frac{1}{3}\delta_{ij}\delta_{kl}$ with the indices $i,j,k,l=1,\ldots,3$ and $a_{d}^{5}(k)=-\text{tan}\delta_{\ell=2}(k)/k^5$ is the energy dependent d-wave scattering length.
The bidirectional operator, $^\leftarrow \!~(\partial_{x_{i}x_{j}}^{2})$ $\left((\partial_{x_{k}x_{l}}^{2})  ^{\rightarrow} \right)$ denotes the differential operator in Cartesian coordinates,
that acts to the left (right) of the pseudopotential $V_{ps,d}(\mathbf{r})$.

We emphasize that application of the expressions (\ref{6}) and (\ref{7}) simultaneously is inconsistent and incompatible 
since the pseudopotentials $V_{ps,s}(\mathbf{r})$ and $V_{ps,d}(\mathbf{r})$ are valid for the cases where the wavefunction exhibits only $1/r$- or $1/r^3$- singularities for small $r$, respectively.

\subsection{Hamiltonian, scattering states and virtual excitations}
We assume two-body collisions of identical bosons in a waveguide with their transversal degrees of freedom being confined by a harmonic oscillator potential.
The harmonic confinement permits the separation of the center of mass and relative motion yielding
a single-particle Hamiltonian with a scatterer fixed at the origin for the relative degree of freedom.

\begin{align}\nonumber
H(z,\rho,\phi) = -\frac{\hbar^2}{2\mu} \left(\frac{\partial ^{2}}{\partial z^{2}}+\frac{\partial ^{2}}{\partial \rho^{2}}+\frac{1}{\rho}\frac{\partial}{\partial \rho}
+\frac{1}{\rho^{2}}\frac{\partial ^{2}}{\partial \phi^{2}}\right )\\
+\frac{1}{2}\mu \omega_{\perp}^{2} \rho^{2}+V_{ps,\ell}(\bf r),
\label{8}
\end{align}

\noindent where $\bf r= \bf r_{1}-\bf r_{2}$ is the relative coordinate of the two bosons, 
and $V_{ps,\ell}(\bf r)$ is the pseudopotential which models the arbitrary $\ell$-wave inter-atomic interaction.

The symmetry of the transversal potential, $\frac{1}{2}\mu \omega_{\perp}^{2} \rho^{2}$, implies that the scattering solutions should be expanded in an axially symmetric basis.
Such a basis set are the eigenstates of the two-dimensional (2D) harmonic oscillator which satisfy the Schr\"{o}dinger equation

\begin{align}\nonumber
\left[-\frac{\hbar^2}{2\mu} \left(\frac{\partial ^{2}}{\partial \rho^{2}}+\frac{1}{\rho}\frac{\partial}{\partial \rho}+\frac{1}{\rho^{2}}
\frac{\partial ^{2}}{\partial \phi^{2}} \right) + \frac{1}{2} \mu \omega_{\perp}^{2} \rho^{2}\right]\tilde{\varPhi}_{n,m_z}(\rho,\phi)\\
=E_{n,m_{z}}\tilde{\varPhi}_{n,m_z}(\rho,\phi),
\normalsize{\label{9}}
\end{align}

\noindent where $\tilde{\varPhi}_{n,m_z}(\rho,\phi)$ are the eigenfunctions of the 2D harmonic oscillator, 
$m_{z}$ is the azimuthal quantum number and $E_{n,m_{z}}=\hbar\omega_{\perp}(n+\lvert m_{z}\rvert+1)$
is the transversal energy spectrum of the 2D harmonic oscillator, with $n=0,2,4,\ldots$ being the principal quantum number.

The spherical symmetry imprinted by the pseudopotential, $V_{ps,\ell}(\bf r)$, leads to a separation of the azimuthal $\phi$ angle from the $z$, $\rho$ coordinates in Eq.(\ref{9}).
The latter means that during the collision no virtual excitations will occur with respect to the $m_{z}$ quantum number.
However, virtual excitations will emerge with respect to the principal quantum number $n$, which will be taken into account in the calculation given below.

We focus on the single mode regime, where the collision takes place in the ground state of the transversal confinement with angular momentum $m_{z}=0$ and represents the only ``open channel''.
The latter means that the total energy of the pair collision is limited between the ground and the first axially symmetric transversal state

\begin{equation}
E_{n=0,m_{z}=0}\leqslant E < E_{n=2,m_{z}=0}.
\label{10}
\end{equation}

Furthermore, the above-mentioned independence on the angle $\phi$ results in the following simplified Schr\"{o}dinger equation:

\begin{align}\nonumber
\left[-\frac{\hbar^2}{2\mu} \left ( \frac{\partial ^{2}}{\partial z^{2}}+\frac{\partial ^{2}}{\partial \rho^{2}}
+\frac{1}{\rho}\frac{\partial}{\partial \rho}\right )+\frac{1}{2} \mu \omega_{\perp}^{2} \rho^{2}+V_{ps,\ell}(\bf r)\right]\Psi(z,\rho)\\
=E \Psi(z,\rho),
\label{11}
\end{align}

\noindent where $E=E_{\parallel}+E_{n=0,m_{z}=0}$ is the total colliding energy, which is defined as the sum of the longitudinal kinetic energy $E_{\parallel}=\dfrac{\hbar^{2}}{2\mu}k^{2}$  and the energy of the
transverse ground state $E_{n=0,m_{z}=0}=\hbar\omega_{\perp}$. Due to the constraint of Eq.(\ref{10}) the momentum $k$ is limited according to the relation
$k<\sqrt{\dfrac{2\mu}{\hbar^{2}}\left(E_{n=2,m_{z}=0}-E_{n=0,m_{z}=0}\right)}$. Here, $\Psi(z,\rho)$ denotes the full 2D axially symmetric solution, which can be expanded in 
the $ \tilde{\varPhi}_{n,0}(\rho,\phi) \equiv \varPhi_n(\rho)$ basis

\begin{equation}
\Psi(z,\rho)=\sum_{n=0}^\infty  C_{n}(z)\varPhi_{n}(\rho)
\label{12}
\end{equation}
where we sum over all even $n$ due to the bosonic symmetry.

Substituting Eq.(\ref{12}) into Eq.(\ref{11}), multiplying by $2\pi\rho\varPhi_{n}^{\ast}(\rho)$, integrating and using Eq.(\ref{9}) yields the following equation for the functions $C_{n}(z)$:

\begin{align}\nonumber
\left[-\frac{\hbar^2}{2\mu}\frac{\partial ^{2}}{\partial z^{2}}+E_{n,m_{z}=0}-E\right] C_{n}(z)= ~~~~~~~~~~~~~~~~~~~~~~\\
-\int_{0}^\infty \varPhi_{n}^{\ast}(\rho)V_{ps,\ell}(\mathbf r)~\Psi(z,\rho)~2\pi\rho d\rho.
\label{13}
\end{align}

By solving Eq.(\ref{13}) for $n=0$ we obtain the scattering wavefunction $C_{0}(z)$ in the open channel which has the following asymptotic form:

\begin{equation}
C_{0}(z) = \cos (kz)+f_{e,\ell}e^{ik\lvert z\rvert}, ~~ \text{for}~\lvert z\rvert\rightarrow \infty,
\label{14}
\end{equation}

\noindent where $f_{e,\ell}$ is the one-dimensional scattering amplitude of the $\ell$-wave interaction which describe the even scattered waves.

For $n>0$, Eq.(\ref{13}) provides us with the solution for the virtual excitations associated with the closed channels, which decay exponentially according to the relation

\begin{equation}
C_{n}(z)=A_{n,\ell}e^{-\sqrt{\frac{n}{2}-\left(\frac{ka_{\perp}}{2}\right)^{2}} \frac{2\lvert z\rvert}{a_{\perp}}},
\label{15}
\end{equation}

\noindent where the coefficients $A_{n,\ell}$ refer to the $\ell$-wave interaction and denote the transition amplitudes
from the transverse ground state to the $n$-th excited state.

In the case of the d-wave pseudopotential we substitute Eq.(\ref{14}) into Eq.(\ref{13}) for $n=0$ and Eq.(\ref{15}) into Eq.(\ref{13}) for $n>0$ and integrate over the $z$ variable , respectively, 
in the interval $[-\epsilon,\epsilon]$ with $\epsilon\rightarrow0$. 
Finally, we obtain the following relations for the d-wave scattering and transition amplitudes:

\begin{align}
f_{e, d}=i\frac{a_{d}^5(k) \sqrt{\pi}}{24ka_{\perp}^3}\eta_d\text~{;}~~
~~A_{n, d}=\frac{a_{d}^5(k) \sqrt{\pi}}{48a_{\perp}^3}\eta_d\frac{2n+1}{\sqrt{\frac{n}{2}-\left(\frac{ka_{\perp}}{2}\right)^{2}}},
\label{17}
\end{align}

\noindent where $\eta_d$ is the regularized part of the wavefunction $\Psi$ in the limit $\mathbf{r}\rightarrow 0$,

\begin{align}\nonumber
\eta_d=~~~~~~~~~~~~~~~~~~~~~~~~~~~~~~~~~~~~~~~~~~~~~~~~~~~~~~~~~~~~~~~~~~~~~\\
\frac{\partial ^5}{\partial z^5} \Bigg\{ z^5 \left[(\partial_{x}^2 +\partial_{y}^2) \Psi(z,\rho)\Big| _{{x\rightarrow0} \atop y \rightarrow0}-2 \partial_{z}^2 \Psi(z,\rho)
\Big| _{{x\rightarrow0} \atop y \rightarrow0}\right]\Bigg\}
\Bigg| _{z \rightarrow0^+},
\label{18}
\end{align}
with $\rho$ given by the relation $\rho=\sqrt{x^2+y^2}$.

Then the expression for the wavefunction reads

\begin{align}\nonumber
\Psi(z,\rho)=\left[ \cos(kz)+i\frac{a_{d}^5(k) \sqrt{\pi}}{24ka_{\perp}^3}\eta_de^{ik\lvert z\rvert}\right]\varPhi_{0}(\rho)\\
+\frac{a_{d}^5(k) \sqrt{\pi}}{48a_{\perp}^2}\eta_d \Lambda_1(z,\rho),
\label{19}
\end{align}

\noindent where $\Lambda_1(z,\rho)$ is defined as follows:

\begin{equation}
 \Lambda_1(z,\rho)=
\sum_{n=2}^\infty \frac{2n+1}{\sqrt{\frac{n}{2}-\left(\frac{ka_{\perp}}{2}\right)^{2}}} ~e^{-\sqrt{\frac{n}{2}-\left(\frac{ka_{\perp}}{2}\right)^{2}}
\frac{2 \lvert z\rvert}{ a_{\perp}}} \varPhi_{n}(\rho).
\label{20}
\end{equation}

Let us now to proceed with the explicit definition of the regular part $\eta_d$ of the wavefunction, being a solution of Eq.(\ref{18}).
The action of the operator $\left(\partial_{x}^2 +\partial_{y}^2\right)$ on $\Psi (\mathbf{r})$ for $ x,y \rightarrow0$ results in the following relation:

\begin{align}\nonumber
\left(\partial_{x}^2 +\partial_{y}^2\right)\Psi(z,\rho)\Big| _{{x\rightarrow0} \atop y \rightarrow0}=~~~~~~~~~~~~~~~~~~~~~~~~~~~~~~~~~~~~~~~~~~~~~~~~\\
-\frac{2}{a_{\perp}^3\sqrt{\pi}}\left[ \cos(kz)+i\frac{a_{d}^5(k) \sqrt{\pi}}{24ka_{\perp}^3}\eta_de^{ik\lvert z\rvert}\right]
-\frac{a_{d}^5(k)}{24a_{\perp}^5}\eta_d \Lambda_2(z,\rho=0),
\label{21}
\end{align}

\noindent where $\Lambda_2(z,\rho=0)$ reads:
\begin{equation}
 \Lambda_2(z,\rho=0)=
\sum_{n=2}^\infty \frac{(2n+1)^2}{\sqrt{\frac{n}{2}-\left(\frac{ka_{\perp}}{2}\right)^{2}}} ~e^{-\sqrt{\frac{n}{2}-\left(\frac{ka_{\perp}}{2}\right)^{2}}
\frac{2 \lvert z\rvert}{ a_{\perp}}}.
\label{22}
\end{equation}

Here, we should note that the $\Lambda$-series at $\rho=0$ are not converging uniformly as $z\rightarrow0$.
The latter means that the partial derivative w.r.t. $z$-coordinate cannot be interchanged with the sum over all even $n$.
Thus, firstly one has to perform the summation and then differentiate with respect to the $z$ coordinate.
However, with the help of Euler's summation formula \cite{apostol99} one can express the $\Lambda$-series as an expansion with respect to the $z$ variable,
where the singular part of the $\Lambda$-series will be written in closed form yielding the following expressions:

\begin{align}
 \Lambda_1(z,\rho=0)&=\frac{2a^3_{\perp}}{\lvert z\rvert^3}+\frac{a_{\perp}}{\lvert z\rvert}+q_1+q_2\frac{2\lvert z\rvert}{a_{\perp}}+q_3\left(\frac{2\lvert z\rvert}{a_{\perp}}\right)^2+\cdots,\\
 \Lambda_2(z,\rho=0)&=\frac{24a^5_{\perp}}{\lvert z\rvert^5}+\frac{4a^3_{\perp}}{\lvert z\rvert^3}+\frac{a_{\perp}}{\lvert z\rvert}+b_1+\cdots,
\label{24}
\end{align}

\noindent where $q_1\approx-2.29225$, $q_2=0.83333$, $q_3\approx-0.154912$ and $b_1\approx-3.5336$ are constants calculated via Euler's summation formula.

Consequently, by substituting Eqs.(\ref{19})-(\ref{24}) into Eq.(\ref{18}) we obtain $\eta_d$, which is now free of divergences.
This in turn yields the following expression for the one-dimensional scattering amplitude:

\begin{equation}
 f_{e,d}=-\frac{1}{1+i\left( \frac{ka_{\perp}}{1-(ka_{\perp})^2}\right) \left(- \frac{a_{\perp}^5}{10a_d^5(k)}+b_2\right)},
\label{25}
\end{equation}

\noindent where the constant $b_2\approx 2.386445$ is defined by the relation $b_2=-\frac{b_1}{2}-4q_3$. The resonance condition for $f_{e,d}$ then reads

\begin{equation}
  \frac{a_{\perp}}{a_d(k)}=\sqrt[5]{10b_2}.
\label{26}
\end{equation}

Eq.(\ref{25}) shows that the d-wave scattering in the presence of the waveguide becomes resonant at an off-resonant value of the free space $a_d(k)$ scattering length, 
due to the significant contributions of the virtual excitations.
The latter holds equally for the case of s-wave CIR. 
Therefore, one can conclude that in general the appearance of resonances in systems with transversal confinement are generated by the virtual excitations in the transverse modes.

By applying the s-wave pseudopotential, given in Eq.(\ref{6}), and performing the derivation as mentioned above,
we will obtain the same relation for the one-dimensional scattering amplitude as in \cite{olshanii98}
with the difference that in this case the s-wave scattering length is energy dependent \cite{tiesinga00}:

\begin{equation}
 f_{e,s}=-\frac{1}{1+ika_{\perp} \left(- \frac{a_{\perp}}{2a_s(k)}+\frac{c_1}{2}\right)},
\label{27}
\end{equation}
The constant $c_1$ in Eq.(\ref{27}) is equal to $1.4603$.

\section{Results and discussion}

In this section we compare the analytical results of the previous subsection for the
s- and d-wave pseudopotentials with the corresponding numerical simulations.
First we evaluate the d- and s-wave energy dependent
scattering lengths by solving numerically a model of Cs atoms in
free space interacting via a Lennard-Jones potential,
$V(r)=\frac{C_{12}}{r^{12}}-\frac{C_6}{r^6}$. 
Note that Cs atoms are ideal canditate for studing scattering phenomena, since they provide a rich spectrum of resonances \cite{chin04, mark07}.

The dispersion
coefficient $C_6$ has been taken from \cite{derevianko01,derevianko02} and $C_{12}$ is a
free parameter, which controls the values for the s- and d-wave scattering lengths.
This yields a parametrization of the one-dimensional s- and d-wave scattering amplitudes, Eqs.(\ref{27})
and (\ref{25}), respectively, in terms of $C_{12}$. 

In order to numerically solve the Cs-Cs collisions in the harmonic waveguide we employ the units $m_{\text{Cs}}/2 = \hbar = \omega_0 = 1$, where $m_{\text{Cs}}$ is 
the mass of the Cs atom and $\omega_0 = 2\pi\times 10$ MHz.
The longitudinal energy is set to $\varepsilon_{\parallel} = 2 \times 10^{-6}$ and the transversal energy is varied within the interval $ 10^{-5} \leq \varepsilon_{\perp} \leq 8 \times 10^{-4}$,
corresponding to a range $0.2\pi~\text{kHz} \leq \omega_{\perp} \leq 16\pi~\text{kHz}$ for the waveguide confinement frequency.
We thereby focus on the low energy regime, characterized by $ka_{\perp} \ll 1$.
In the following we calculate the transmission coefficient given by:
\begin{equation}
 T_\ell=\lvert1+f_{e,\ell}\rvert^2,
\label{28}
\end{equation}
where $f_{e, \ell}$ is the one-dimensional scattering amplitude for the $\ell$-wave.

\begin{figure}[!t]
\includegraphics[width=0.7\columnwidth]{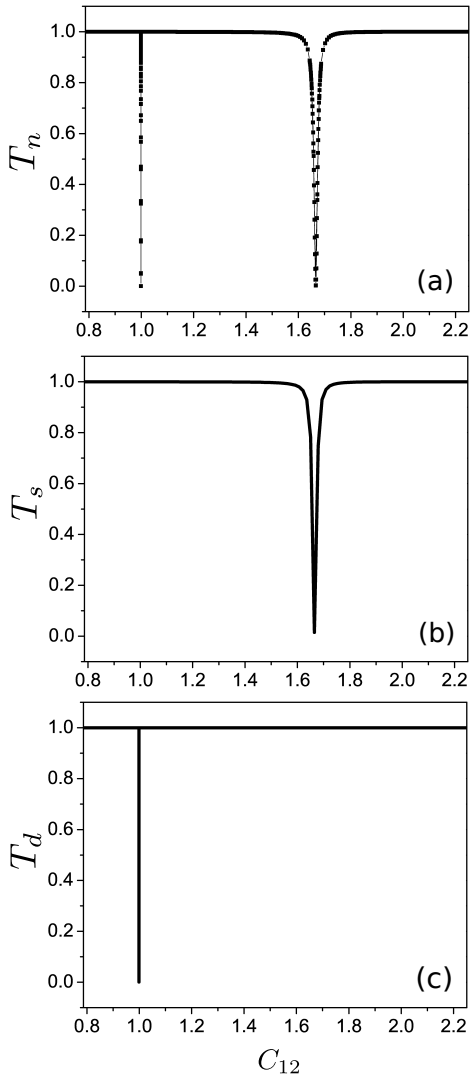}
\caption{(a) Numerically calculated transmission coefficient $T_{n}$,which shows the DWR on the l.h.s and the s-wave CIR on the r.h.s..
The analytically calculated (b) s-wave transmission coefficient, $T_s$, and (c) d-wave transmission coefficient, $T_d$.
The transversal confinement is $\omega_{\perp}=10^{-5}$}.
\label{fig1}
\end{figure}

In Fig.~\ref{fig1} we show the transmission coefficient as a function of 
$C_{12}$ with the transversal frequency $\omega_{\perp}=10^{-5}$.
$T_{n}$ in Fig.~\ref{fig1}(a) denotes the numerically calculated transmission coefficient, as it was shown in \cite{gian11, melezhik91, saeidan08}. 
We observe the appearance of two minima ($T_{n}=0$) due to resonances.
The minimum on the r.h.s. of Fig.~\ref{fig1}(a) refers to the s-wave CIR and the minimum on the l.h.s. to the DWR.
Due to the centrifugal barrier the width of the DWR is much narrower than that of the s-wave CIR.
Additionally, we observe that both resonances yield a Lorentzian-like lineshape in the transmission spectrum.
This occurs due to the fact that the weak confinement suppresses the interference effects between the s-wave and d-wave interactions.

\begin{figure}[hbt]
\includegraphics[width=1.0\columnwidth]{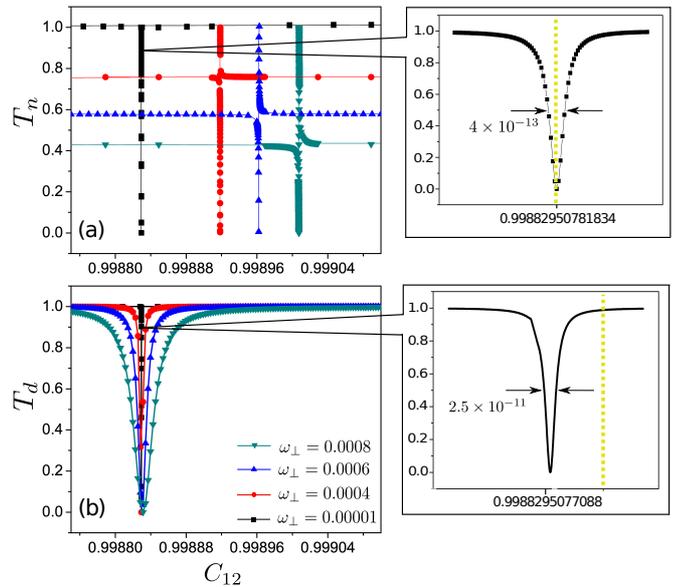}
\caption{(color online)(a)(left panel) The numerically calculated transmission coefficient $T_{n}$ versus $C_{12}$ parameter for several confinement frequencies $\omega_{\perp}$, 
(right panel) high-resolution figure of the DWR for $\omega_{\perp}=10^{-5}$ where the inwards-pointing arrows denote the width of the resonance.
(b)(left panel) The analytically calculated transmission coefficient $T_d$ as a function of the $C_{12}$ parameter calculated with the d-wave pseudopotential for the corresponding values of $\omega_{\perp}$,
(right panel) high-resolution graph of the DWR for $\omega_{\perp}=10^{-5}$.
The yellow dashed line in the right panels indicates the minimum of the numerical calculated transmission coefficient ($T_{n}=0$) in comparison with the corresponding analytical result.}
\label{fig2}
\end{figure}

In Fig.~\ref{fig1}(b) the analytically calculated transmission coefficient $T_s$, as given by Eqs.(\ref{27}) and (\ref{28}), is shown which refers to s-wave interactions only.
$T_s$ describes very accurately the s-wave CIR (see Fig. \ref{fig1}(a)-(b)), whereas there is no trace of the DWR,
since contributions from higher partial waves are not included in Eq.(\ref{27}).
In Fig.~\ref{fig1}(c) the analytic expression of the transmission coefficient $T_d$ is depicted according to Eqs.(\ref{25}) and (\ref{28}).
We observe that the DWR is described qualitatively accurate enough (see also insets of Figs.\ref{fig2}(a)-(b)), despite the fact that in the analytical calculations we have neglected the contributions of s-wave scattering,
since the transversal confinement is very weak.
We note that the results shown in Figs.~\ref{fig1}(b) and \ref{fig1}(c) being obtained analytically for the pure s- and d-wave scattering cases
lead to an accurate description of the scattering processes in the presence of a weak external harmonic potential.

Let us now determine the regime of validity of the approximative analytical results for the d-wave pseudopotential scattering by comparing it with the corresponding numerical simulations.
In Fig.~\ref{fig2} we show the transmission coefficient $T$ calculated numerically (Fig.~\ref{fig2}(a)) and analytically (Fig.~\ref{fig2}(b)) for several transversal confinement frequencies $\omega_{\perp}$.
In Fig.~\ref{fig2}(a) for a strong confinement, e.g. $\omega_{\perp}=8\times10^{-4}$, we first observe that the numerically calculated position of the resonance deviates substantially from the analytical one.
However, as $\omega_{\perp}$ decreases the numerical results for the position of the DWR converges to the corresponding analytical one.
The origin of this behavior is the fact that we considered only d-wave interactions in the analytical calculations and neglected contributions from the s-wave interactions.
This approximation is eligible in the regime of a weak confinement, e.g. $\omega_{\perp}\leqslant4\times10^{-5}$, where the s- and d-wave free space resonances are weakly coupled,
and in the vicinity of the DWR the d-wave interactions are dominant over the s-wave.
In the regime of strong confinement the transmission spectrum $T$, in the vicinity of the DWR, exhibits a strong asymmetric Fano-lineshape,
with $T_{n}$ changing abruptly from zero to one. The latter is an interference effect of the strong s- and d-wave interactions in the confinement and cannot be described by this analytical pure d-wave model.
However, this asymmetric profile gets suppressed with decreasing $\omega_{\perp}$ yielding a Lorentzian-like lineshape.

The right panels of Fig.~\ref{fig2} show a high-resolution graph of the DWR calculated analytically and numerically for $\omega_{\perp}=10^{-5}$,
where the dashed vertical line in both plots indicates the position of the minimum of the numerically calculated transmission coefficient $T_{n}$.
We observe that the d-wave pseudopotential approximation qualitatively describes the numerical simulations in the regime of weak confinement.
However, in this small scale of the $C_{12}$ values the analytical results show that the width of the DWR is by two orders wider than the numerical predictions,
 as well as there is a slight difference in the position of the DWR.
These quantitative deviations again occur due to s-wave interactions,
which act as corrections in the transmission coefficient $T_{n}$ despite of the weak coupling of the s-wave  and d-wave interactions.

\section{Brief Summary}

We have derived and analyzed analytical expressions for resonant
d-wave scattering in a harmonic waveguide, in the framework of
pseudopotential theory. The interatomic
interaction has been modeled by a pure d-wave pseudopotential introduced in
\cite{idziaszek06,idziaszek09}. We observe that the analytical
pseudopotential approach provides results which are in qualitative agreement with the corresponding numerical results in the regime of weak confinement due to the very small
interference effects of s- and d-wave interactions.The appearance of the DWR in a waveguide is strongly affected by the virtual excitations over all the closed excited modes of the transversal confinement. 
For a strong transversal confinement the analytically and numerically calculated lineshape and position of the DWR deviate from each other.
The interference effect of the s- and d-wave free space resonances in the confinement yields an asymmetric Fano profile which cannot be derived by a single partial wave model. 
Our analytical result Eq.(\ref{25}) for weak confinement
can be used for the analysis of resonant
scattering processes in confined geometries beyond s-wave
physics. 
Finally, we remark that our work clearly demonstrates the necessity of an extended   
pseudopotential theory capable of describing systems which 
possess a non-spherical symmetry that yield interference phenomena
between different partial wave scattering amplitudes.

\begin{acknowledgements}
P.G. thanks C. Morfonios and F.K. Diakonos for valuable comments and discussions.
V.S.M. acknowledges financial support by the Deutsche Forschungsgemeinschaft and the Heisenberg-Landau Program. P.S. acknowledges the Deutsche Forschungsgemeinschaft for financial support. 

\end{acknowledgements}

\end{document}